\documentclass[mathpazo]{cicp}
\usepackage{generic}
\usepackage{cite}
\usepackage{amsmath,amssymb,amsfonts}
\usepackage{algorithmic}
\usepackage{graphicx}
\usepackage{float}
\usepackage{textcomp}
\usepackage{multirow}
\usepackage{color}
\usepackage{hyperref}

\begin{document}
\title{Fast MRI Reconstruction via Edge Attention}


\author[O.~Author]{Hanhui Yang\affil{1}, Juncheng Li\affil{2}\corrauth, Lok Ming Lui\affil{1}, Shihui Ying\affil{3}, Jun Shi\affil{2}, and Tieyong Zeng\affil{1}*}
\address{
    \affilnum{1} The Department of Mathematics, The Chinese University of Hong Kong, New Territories, Hong Kong \\
    \affilnum{2} The School of Communication and Information Engineering, Shanghai University, Shanghai, China \\
    \affilnum{3} The Department of Mathematics, School of Science, Shanghai University, China}
\emails{{\tt huihuibullet43@gmail.com} (H.~Yang), {\tt cvjunchengli@gmail.com} (J.~Li), {\tt lmlui@math.cuhk.edu.hk} (L.~Lui), {\tt shying@shu.edu.cn} (S.~Ying), {\tt junshi@shu.edu.cn} (J.~Shi), {\tt zeng@math.cuhk.edu.hk} (T.~Zeng)}


\begin{abstract}
Fast and accurate MRI reconstruction is a key concern in modern clinical practice. Recently, numerous Deep-Learning methods have been proposed for MRI reconstruction, however, they usually fail to reconstruct sharp details from the sub-sampled $k$-space data. To solve this problem, we propose a lightweight and accurate Edge Attention MRI Reconstruction Network (EAMRI) to reconstruct images with edge guidance. Specifically, we design an efficient Edge Prediction Network to directly predict accurate edges from the blurred image. Meanwhile, we propose a novel Edge Attention Module (EAM) to guide the image reconstruction utilizing the extracted edge priors, as inspired by the popular self-attention mechanism.
EAM first projects the input image and edges into $\mathbf{Q_{image}}$, $\mathbf{K_{edge}}$, and $\mathbf{V_{image}}$, respectively. Then EAM pairs the $\mathbf{Q_{image}}$ with $\mathbf{K_{edge}}$ along the channel dimension, such that 1) it can search globally for the high-frequency image features that are activated by the edge priors; 2) the overall computation burdens are largely reduced compared with the traditional spatial-wise attention.
With the help of EAM, the predicted edge priors can effectively guide the model to reconstruct high-quality MR images with accurate edges. Extensive experiments show that our proposed EAMRI outperforms other methods with fewer parameters and can recover more accurate edges. 
\end{abstract}

\ams{54H30, 68U03, 68U10, 68T07}
\keywords{Image restoration, MRI reconstruction, Edge attention}

\maketitle

\section{Introduction} \label{intro}
Magnetic Resonance Imaging (MRI) is one of the most important tools in image-guided adaptive radiotherapy, which helps doctors locate pathological regions without harmful radiation exposure. However, due to Nyquist sampling requirement~\cite{nyquist1928certain}, the imaging time is frustratingly long to get high-quality images from fully-sampled $k$-space data. Nowadays, Parallel Imaging (PI) has become a standard strategy used by most clinical MRI scanners to accelerate the imaging process. PI places multiple receiver coils around the subject, each of which subsamples $k$-space data from different views. Therefore, the overall imaging time can theoretically be reduced by a factor of the coil number. 


During the past decades, numerous reconstruction methods have been designed for parallel MR Imaging. Among them, Compressed Sensing (CS) based methods are a broad class of mature and effective methods that are theoretically supported by~\cite{candes2006robust}. CS-based methods exploit the inherent sparsity of MR data in some properly transformed domains and can recover clear images from sub-sampled $k$-space via iteratively solving a constrained optimization problem~\cite{lustig2010spirit, he2022low}. Recently, with its great success in various image processing tasks (e.g., image classification~\cite{he2016deep, tan2019efficientnet} and image restoration~\cite{dong2014learning, zeng2015spectral,CiCP-31-893}), Deep learning (DL) has also greatly promoted the development of parallel MR imaging. For example, Chen \textit{et al.}~\cite{chen2022pyramid} proposed a PC-RNN model with three convolutional RNN (ConvRNN) modules to iteratively learn the features in multiple scales. Aggarwal \textit{et al.}~\cite{aggarwal2018modl} proposed a general model-based image reconstruction framework with a convolution neural network (CNN) based regularization prior. However, existing methods ignore the necessity of edge reconstruction, resulting in a lack of accurate and clear edges in reconstructed MR images.

\begin{figure}[t!]
\centering
\begin{minipage}[c]{0.14\textwidth}
\includegraphics[height=2.7cm]{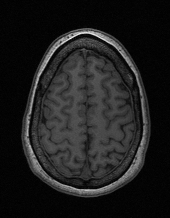}
\includegraphics[height=2.7cm]{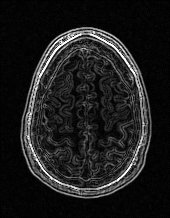}
\centerline{(a)}
\end{minipage}
\begin{minipage}[c]{0.18\textwidth}
\includegraphics[height=2.7cm]{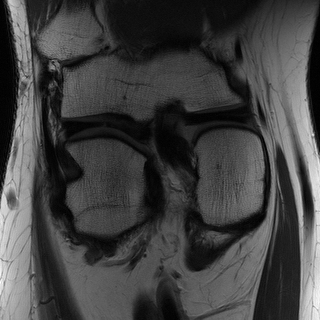}
\includegraphics[height=2.7cm]{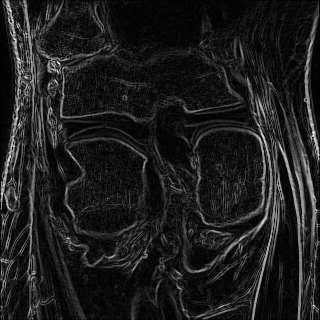}
\centerline{(b)}
\end{minipage}
\begin{minipage}[c]{0.21\textwidth}
\includegraphics[height=2.7cm]{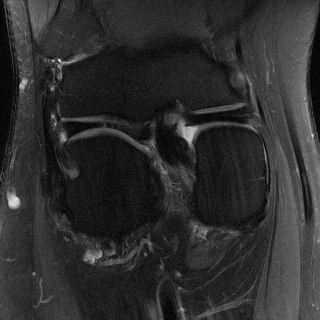}
\includegraphics[height=2.7cm]{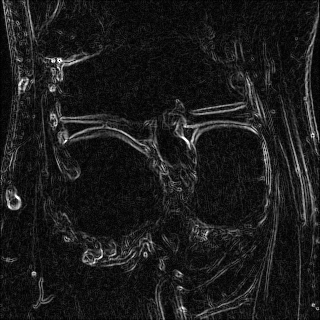}
\centerline{(c)}
\end{minipage}
\caption{Examples of different MR modalities with their edges. (a), (b) and (c) correspond to T1-weighted, PD, and PDFS images, respectively. The upper row are the original images and the second row are the extracted edges with Sobel operator. 
} 
\label{pd-pdfs-diff}
\end{figure}

Edge preserving has always been a crucial concern in the design of reconstruction models. To improve the quality of reconstructed images and preserve image edges, some works suggested introducing edge priors in the original restoration problem to preserve image edges~\cite{belge2000wavelet, ren2013fractional}. However, they will suffer from complicated algorithm design and time-consuming training processes. Recently, some more efficient methods have been proposed to use edge maps as external guidance for image restoration. For example, Yang et al.~\cite{yang2017deep} used off-the-shelf edge detectors to extract image edges from the degraded images. Fang et al.~\cite{fang2020soft} predicted image edges by constructing an edge reconstruction network. Huang~\textit{et al.}~\cite{huang2022edge} designed a novel dual discriminator GAN framework for solving fast multi-channel MRI, in which one GAN network is built for edge information enhancement. Inspired by these methods, we also consider introducing image edge prior as external guidance to MRI reconstruction since 1) image edges are prominent and distinguishable features in MRI (see Fig.~\ref{pd-pdfs-diff}), which can serve as a good guide to the model to recover high-frequency details; 2) the ground truth edges can be easily fetched via ordinary edge extraction operators, like Canny, Sobel, and Prewitt, which means that the edge maps can be learned in a data-driven manner. However, how to effectively utilize image edge priors to guide image reconstruction still remains a challenge. In some methods, edge information was simply concatenated with the input image and passed to the next stages. Though this is a simple way to utilize the edge priors, it may not give full play to the guiding role of the edge priors. Therefore, in this work, we want to explore a more efficient and effective mechanism to fully take advantage of image edge priors.

To address the aforementioned issues, in this work, we propose a lightweight and accurate Edge Attention MRI Reconstruction Network (EAMRI) that utilizes a novel edge attention strategy to ensure the quality of the edges of the reconstructed MR images. Specifically, we design an efficient Edge Prediction Network (EPN) to directly predict accurate image edges from the under-sampled MR image. Meanwhile, we propose a novel Edge Attention Module (EAM) to fully use edge priors to guide image reconstruction. The proposal of EAM is inspired by Transformer, which suggested using the self-attention mechanism to learn the global information in various image processing tasks~\cite{dosovitskiy2020image}. Different from existing Transformers, EAM aims to explore the relationship between image and edge information. In particular, EAM pairs the image queries with the edge keys along the channel dimension, which enables the model to find the high-frequency features that best match edge features. In this way, the network can restore high-frequency MR images under the guidance of edge priors. It is worth noting that different from the self-attention mechanism, whose computational complexity grows quadratically with the spatial resolution, our EAM has linear computational complexity, which is feasible for high-resolution images. In summary, the main contributions of this work are as follows:
\begin{itemize}
    \item An efficient Edge Prediction Network (EPN) is designed to directly predict accurate image edges from the under-sampled MR image. This is the first edge prediction CNN model in the field of MRI reconstruction.
    \item We propose a novel Edge Attention Module (EAM) to fully stimulate the role of edge guidance. With the help of EAM, the predicted edges can effectively guide the model to reconstruct high-quality MR images with sharp and accurate edges.  
    \item We propose a lightweight and accurate Edge Attention MRI Reconstruction Network (EAMRI). Extensive experiments show that EAMRI achieves better results than other SOAT methods with fewer parameters.  
\end{itemize}

The rest of this paper is organized as follows. Related works are reviewed in Section~\ref{sec:related_work}. A detailed explanation of the proposed EAMRI is given in Section~\ref{sec:methods}. The experiment details, results and ablation studies are presented in Section~\ref{sec:experiment} - ~\ref{sec:ablation}, respectively. Finally, we draw a conclusion in Section~\ref{sec:conclusion}.
 
\section{Related Works}\label{sec:related_work}


\subsection{Deep Learning for Parallel MR Imaging}
Parallel Imaging (PI) is one of the most effective techniques to speed up the MRI acquisition process. According to~\cite{knoll2019deep}, traditional PI reconstruction methods can be classified into two types: SENSE-type~\cite{pruessmann1999sense} and GRAPPA-type~\cite{griswold2002generalized} methods. SENSE-type methods usually require the information of sensitivity encoding, and they can eliminate the artifacts in the image domain. Meanwhile, Compress Sensing (CS) theory is often utilized in SENSE-type methods~\cite{knoll2011second,akccakaya2011low,jung2009k}. On the other hand, GRAPPA-type methods eliminate the artifacts by interpolating the missing $k$-space data~\cite{lee2018deep, jung2008parallel}. In recent years, deep learning (DL) has achieved great success in natural image processing, which has also contributed to the development of PI. For example, Huang~\textit{et al.}~\cite{huang2022swin2} proposed the SDAUT for solving fast MRI, which couples Shifted Windows Transformer with U-Net to reduce the network complexity, and incorporates deformable attention to increase the explainability of the reconstruction model. However, GAN-based models may suffer from noise and unstable training~\cite{zhang2021artifact}. On the other hand, a thriving stream of research is derived from the optimization scheme of CS theory applied to the classical SENSE-type methods. For example, Aggarwal~\textit{et al.}~\cite{aggarwal2018modl} proposed an iterative image reconstruction algorithm with a CNN-based regularization prior, which yields a deep network when unrolling. Duan~\textit{et al.}~\cite{duan2019vs} proposed a variable splitting optimization method for generalized parallel CS reconstruction and unrolled it into a trainable network. Interested readers may refer to~\cite{chen2022ai} for a comprehensive review. Some researchers are also interested in applying deep neural networks to learn $k$-space interpolation in a data-driven manner. For instance, Ak{\c{c}}akaya \textit{et al.}~\cite{akccakaya2019scan} proposed to train a 3-layer network on a small amount of scan-specific ACS data and estimated the missing $k$-space data in a non-linear approach. However, these methods are still difficult to reconstruct texture details and edges, which limits their practical application.



\subsection{Edge-guided Image Restoration}
In general, the image restoration (IR) problem can be formulated as an ill-posed inverse problem, which requires regularization (priors) to refine the solution space. During the past decades, researchers have devoted enormous efforts to designing edge-preserving regularization to preserve the sharp edges~\cite{belge2000wavelet, yang2009fast, oektem2017shape}. Among all these methods, total variation (TV) regularization~\cite{rudin1992nonlinear, chen2019new} is one of the most well-known and effective edge-preserving priors. However, these methods usually require hand-crafted designs of regulation terms and may suffer from complex algorithm design. Recently, DL techniques have made great success in various IR tasks, and some works consider using explicit edge priors to guide the reconstruction process. For example, Huang \textit{et al.}~\cite{huang2022edge} proposed the PIDD-GAN with dual GAN discriminator for fast MRI, one for holistic image reconstruction, and the other one for enhancing edge information using the edge information extracted by Sobel operator. Yang \textit{et al.}~\cite{yang2017deep} proposed a recurrent residual network (DEGREE) that utilized the LR image with its corresponding edge maps to infer the sharp details in the HR image. 
However, DEGREE suffers from 1) it extracts edge maps from the learned image features using a simple convolutional operation, which may not be accurate due to the existence of image noises; 2) it produced the final HR outputs by directly adding the predicted edge maps into the LR image, which may not be an efficient way to take full advantage of the edge priors. Considering the above drawbacks, Fang \textit{et al.}~\cite{fang2020soft} introduced an improved edge-guided network (SeaNet), which designed an efficient edge net to predict edge maps and proposed an effective fusion mechanism to combine the edge priors with the learned image features. However, SeaNet still can not give full play to the edge guidance since the edge maps are simply concatenated with the image features. Therefore, we aim to explore an efficient mechanism to fully take advantage of edge guidance.

\section{Methods}\label{sec:methods}
\subsection{Problem Formulation}\label{pf}
In general, the task of parallel MRI reconstruction can be mathematically formulated as:
\begin{equation}
    \mathbf{y_i} = A_i\mathbf{x} + \epsilon_i, \text{ for }  i = 1, 2, \cdots n_c
\label{eq:1}
\end{equation}
where $\mathbf{y_i} \in \mathbb{C}^{n} $ represents the acquired sub-sampled $k$-space from the $i$-th coil and $n_c$ stands for the total number of coils. $\mathbf{x} \in \mathbb{C}^{n}$ is the clear MR image that we want to recover.  $A_i$ is a forward operator with the form of $M \odot F \odot S_i$, where $M$ is a binary mask, $F$ is the Fourier transform operator, and $S_i$ is the coil sensitivity matrix that encodes the spatial sensitivity for each coil. Usually they are normalized as:
\begin{equation}
    \sum_{i=1}^{n_c} S_{i}^{*} S_{i}=1.
\end{equation}
It is known that Eqn.~\ref{eq:1} is an ill-posed inverse problem. Like most DL-based models for MRI reconstruction~\cite{schlemper2017deep, zheng2019cascaded}, we force $\mathbf{x}$ to be well-approximated by our proposed network and solve Eq.~\ref{eq:1} in a supervised learning manner: 
\begin{equation}\label{eq:3} 
\underset{\theta }{\min } \mathcal{L} ( f_{\theta}(\mathbf{x^0}|\mathbf{y}),  \mathbf{x} ), 
\end{equation}
where $\mathbf{y}=\{\mathbf{y_i} \}_{i=1}^{n_c}$ is the acquired measurements from all the coils, $\mathbf{x^0}$ is the zero-filled image converted from $\mathbf{y}$, and $f_{\theta}(\cdot)$ is the proposed EAMRI with learnable parameters $\theta$. More details of the loss function will be covered in Sec.~\ref{loss_func}.



\begin{figure*}[t]
\centering
  \includegraphics[width=\textwidth]{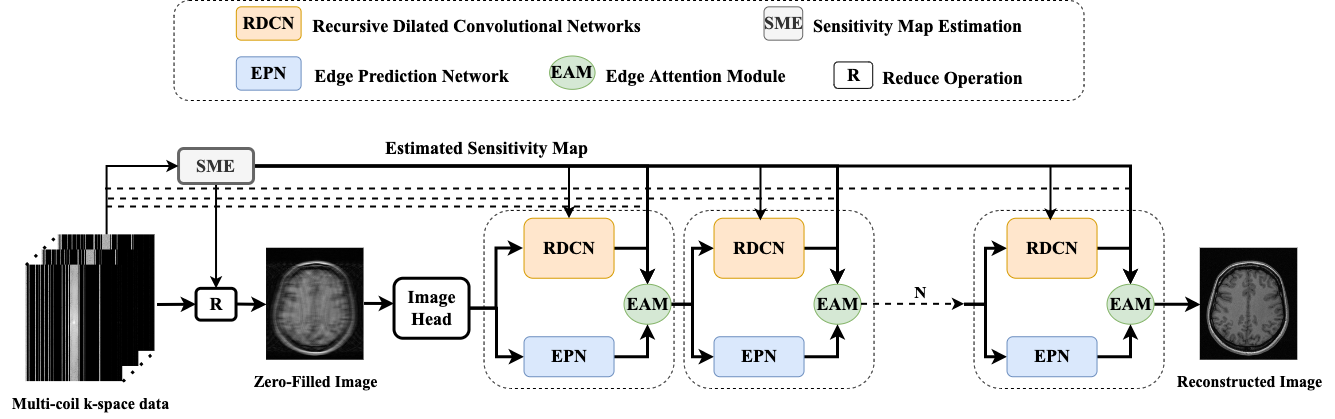}
  \caption{An illustration of the proposed \textbf{EAMRI} framework, which consists of two branches: Image Reconstruction Branch (IRB) and Edge Prediction Branch (EPB). IRB utilizes multiple \textbf{R}ecursive \textbf{D}ilated \textbf{C}onvolutional \textbf{N}etwork (\textbf{RDCN}) for image de-aliasing, while EPB utilizes one recursive component \textbf{E}dge \textbf{P}rediction \textbf{N}etwork (\textbf{EPN}) to extract fine edges from the input image. The predicted edges can guide the network to restore more accurate details with \textbf{E}dge \textbf{A}ttention \textbf{M}odule (\textbf{EAM}).}
  \label{arci} 
\end{figure*}

\subsection{Edge Attention MRI Reconstruction Network}\label{sec:proposed}


In this work, we propose a lightweight and accurate Edge Attention MRI Reconstruction Network (EAMRI). The overall workflow is presented in Fig.~\ref{arci}. Given sub-sampled multi-coil $k$-space data $\mathbf{y}$, we first estimate the sensitivity map $S$ using the Sensitivity Map Estimation (\textbf{SME}) module. The sensitivity map together with the $k$-space data can be used to generate the sensitivity-weighted zero-filled image $\mathbf{x^0}$ via the Reduce operation. Then, we pass $\mathbf{x^0}$ through an Image Head (the same architecture as RDCN, but with fewer parameters), which provides a relatively good initialization for the following part of the network. It is worth noting that the output of Image Head goes through two parallel branches: the image reconstruction branch and the edge prediction branch. The image reconstruction branch consists of multiple Recursive Dilated Convolutional Networks (\textbf{RDCNs}), which are simple but effective image de-aliasing blocks to recover low-frequency image features. On the other hand, the edge prediction branch consists of a recursive Edge Prediction Network (\textbf{EPN}), which can predict accurate image edges from the blurred image. Finally, the predicted edge priors are fed into the Edge Attention Module (\textbf{EAM}), which is an effective attention module that can guide the image details reconstruction with edge priors. More details will be covered in the following sections.

\subsubsection{Sensitivity Map Estimation}\label{sec:sme}
\begin{figure}[t]
\centering
  \includegraphics[width=0.75\textwidth]{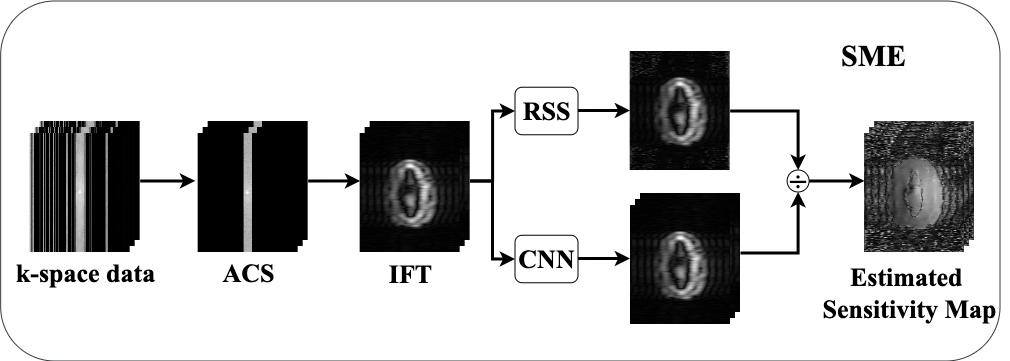}
  \caption{Workflow of the proposed Sensitivity Map Estimation.}
  \label{SME} 
\end{figure}

Inspired by~\cite{sriram2020end}, we introduce the Sensitivity Map Estimation (SME) module to estimate the sensitivity map. As shown in Fig.~\ref{SME}, SME takes the sub-sampled $k$-space data as input and first filters out the Auto-Calibration Signal (ACS) region. Then, the inverse Fourier transform (IFT) is applied on this region to get the blurred images. After that, these images are passed to two parallel branches: 1) RSS, which calculates the root sum of squares of the images; 2) CNN, where a simple CNN is utilized to refine the data. Finally, the sensitivity maps can be estimated by dividing the outputs of these two branches. It is worth noting that different from the widely used sensitivity map estimation method, ESPIRit algorithm~\cite{uecker2014espirit}, in previous works~\cite{hammernik2018learning, aggarwal2018modl}, the proposed SME module is a learnable module that can be jointly trained with the rest parts of the model. For simplicity, the CNN in our SME is chosen to be the same architecture as the following Recursive Dilated Convolutional Network (RDCN) but with fewer parameters.

\subsubsection{Reduce and Expand}
We define two operations in our work, Reduce ($\mathcal{R}$) and Expand ($\mathcal{E}$). As can be seen from Fig.~\ref{arci}, we utilize the Reduce operation at the beginning of the model to provide the initial Zero-Filled image (SENSE reconstruction) for the rest part of the network. Specifically, given sub-sampled $k$-space data $\mathbf{y}=\{\mathbf{y_i} \}_{i=1}^{n_c}$, we first apply the inverse Fourier transform to get image data $\mathbf{\hat{x}} =\{\mathbf{\hat{x}_i} \}_{i=1}^{n_c}$, where each $\mathbf{\hat{x}_i}$ is the individual coil image for $i$-th coil. Then, $\mathbf{x^0}$ can be computed as: 
\begin{equation}\label{reduce}
    \mathbf{x^0} =\mathcal{R} ( S, \mathbf{\hat{x}})  =  \sum_{i=1}^{n_c} S^{*}_i\mathbf{\hat{x}_{i}},
\end{equation}
where $S_i$ the sensitivity map of the $i$-th coil.

On the other hand, we employ the Expand operation in the later discussed Data Consistency (DC) layer. Expand is the inversion the Reduce, which can convert the SENSE reconstruction (for simplicity, we abuse the notation $\mathbf{x^0}$ here) back into individual coil images:
\begin{equation}
    \mathcal{E} (\mathbf{x^0}) = (S_{1}\mathbf{x^0}, S_{2}\mathbf{x^0}, \cdots, S_{n_c}\mathbf{x^0}).
\end{equation}

\subsubsection{Recursive Dilated Convolutional Network}

\begin{figure}[h!]
\centering
  \includegraphics[width=0.57\textwidth]{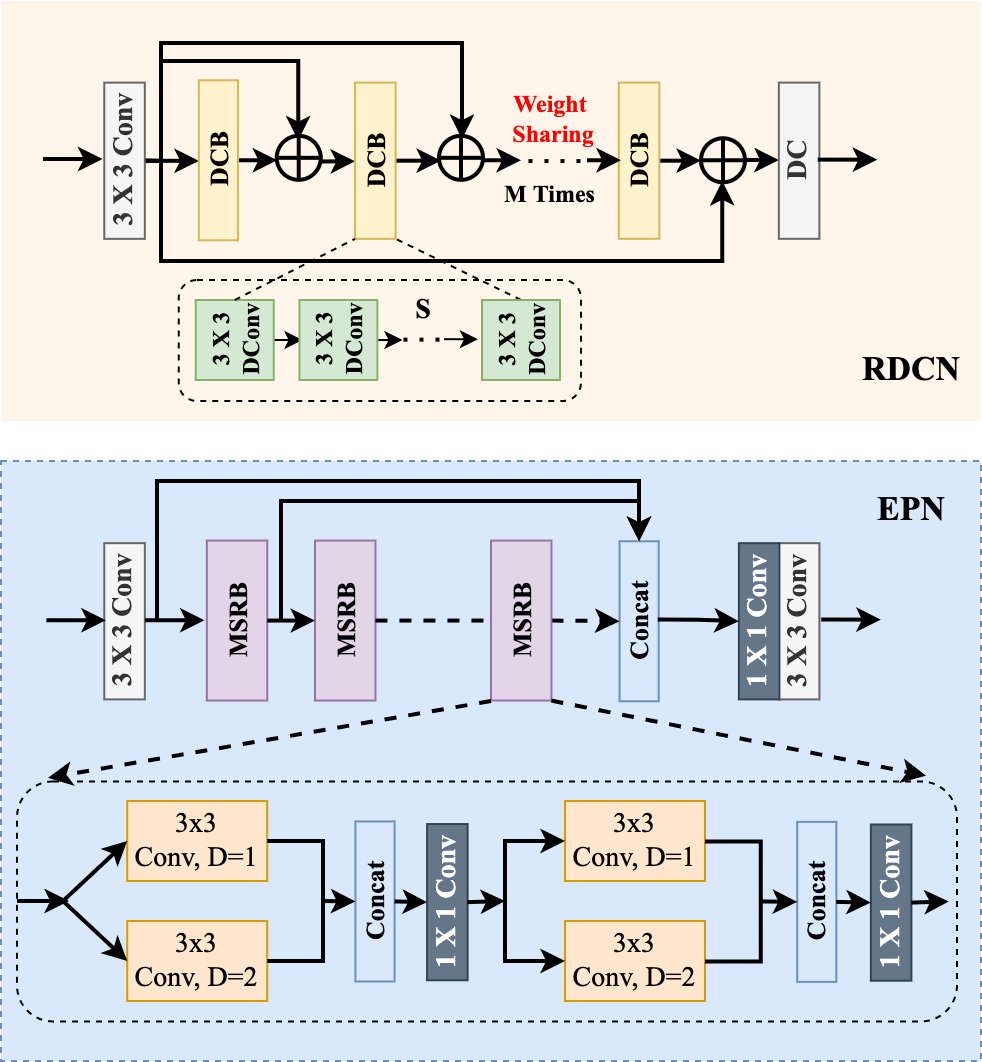}
  \caption{Upper: the proposed Recursive Dilated Convolutional Network (RDCN). The RDCN consists of multiple Dilated Convolution Blocks (DCB), which are \textcolor{red}{weight sharing}; Lower: the proposed edge Prediction Network (EPN). In EPN, we utilize dilated convolution to increase the receptive fields.}
  \label{RDB} 
\end{figure}

Following~\cite{sun2018compressed}, we propose a Recursive Dilated Convolutional Network (RDCN) as the building block for image de-aliasing in the image reconstruction branch. RDCN adopts two design ideas: 1) Recursive learning. As shown in Fig.~\ref{RDB} (Upper), RDCN consists of a repeating component Dilated Convolution Block (DCB) with sharing weights. This is beneficial for the reuse of model parameters and lightweight model design. In the meantime, skip connections are added between each recurrent so that features from the shallow layers can be aggregated in the deeper layers; 2) Dilated convolution. Within each DCB, there are multiple $3 \times 3$ convolutional kernels with different dilation rates, which is beneficial for learning multi-scale features since dilation kernels can effectively increase the receptive fields without bringing additional parameters. From this point of view, RDCN is a lightweight and effective image de-aliasing block. Meanwhile, to maintain data consistency, we also add a DC layer at the bottom of RDCN. 

\subsubsection{Edge Prediction Network}
To obtain accurate edge priors, we propose an efficient Edge Prediction Network (EPN) to directly predict image edges from the input image. It is worth noting that we proposed a similar edge net with~\cite{fang2020soft}, which was used to extract image edges from natural images. In contrast, the EPN in our work is applied to the reconstructed sensitivity-weighted images. As shown in Fig.~\ref{RDB} (Lower), we first use a $3 \times 3$ convolutional layer to increase the feature channels. Then, a cascaded of Multi-scale Residual Blocks (MSRB) is utilized to extract image features at different scales, which is beneficial to capture high-frequency features in the image. Specifically, each MSRB consists of two convolution branches to operate features with different receptive fields. Different from~\cite{li2018multi}, we replace the original $5 \times 5$ convolutional layers with the $3 \times 3$ convolutional layers with dilation rate 2. This change will make the module to has the same receptive field but with fewer parameters. Finally, the outputs of all MSRBs are concatenated and fused by $1 \times 1$ convolutional layer, and a $3 \times 3$ convolutional layer is used to decrease the feature channels to obtain the predict edge maps.

\subsubsection{Edge Attention Module}

\begin{figure}[h!]
\centering
  \includegraphics[width=0.7\textwidth]{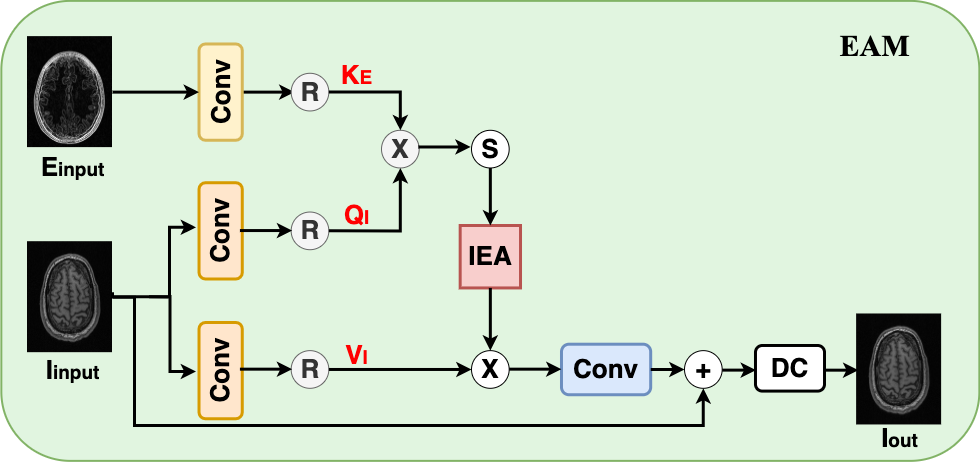}
  \caption{The proposed Edge Attention Module \textbf{(EAM)} to fuse the information of $\operatorname{E_{input}}$ and $\operatorname{I_{input}}$. Here \textbf{R} stands for the reshape operation, and \textbf{S} stands for the Softmax operation.}
  \label{EAM} 
\end{figure}
The core idea of this paper is to use edge priors to guide the model to reconstruct accurate MR images with clean and accurate edges. Most of the existing methods use the concatenation operation to directly combine image features and edge features directly. However, those methods usually fail to fully exploit the role of edge guidance and thus cannot fully utilize edge priors. To solve this issue, we propose a novel Edge Attention Module (EAM), which is an efficient attention module that can utilize the predicted image edges to guide the model to reconstruct high-quality images with more accurate details. It is worth mentioning that, as far as we know, this is the first attempt to utilize edge attention in the field of accelerated MRI reconstruction. As shown in Fig.~\ref{SME}, EAM takes the output ($\operatorname{I_{input}}$) of RDCN and the output ($\operatorname{E_{input}}$) of EPN as inputs. For the image input $\operatorname{I_{input}}$, we generate the corresponding image queries ($\mathbf{Q_{I}}$) and image values ($\mathbf{V_{I}}$) by first applying $1 \times 1$ convolutions to aggregate the channel information plus $3 \times 3$ depth-wise convolutions to encode the channel-wise spatial context. Similarly, we generate the edge keys ($\mathbf{K_{E}}$) by applying $3 \times 3$ convolutions to project the edges into the same dimension space as the queries. Therefore, $\mathbf{Q_{I}}$, $\mathbf{K_{E}}$, and $\mathbf{V_{I}}$ have the same shape as $H \times W \times C$. Then, we reshape $\mathbf{Q_{I}}$ and $\mathbf{K_{E}}$ to $C \times HW$, such that their dot-product interaction generates the Image Edge Attention (\textbf{IEA}) of size $C \times C$. It is worth noting that each element in IEA represents the channel-wise attention weight between $\mathbf{Q_{I}}$ and $\mathbf{K_{E}}$. This novel design distinguishes us from the self-attention module in that we pair $\mathbf{Q_{I}}$ and $\mathbf{K_{E}}$ channel-wisely ($C \times C$) instead of spatial-wisely ($HW \times HW$). In other words, EAM can search for the best match between the image queries and the edge keys along the channel dimension. In this way, we not only can attend to the high-frequency features in a global context manner, but we also save from cumbersome computation of regular attention map. To maintain data consistency, we also add a DC layer at the end. Overall, the EAM process can be defined as:
\begin{equation}
    \begin{aligned}
        &\operatorname{I_{res}} = V_{I} \cdot \operatorname{Softmax}(K_{E} \cdot Q_{I}) / \alpha, \\
        & \operatorname{I_{out}} = \operatorname{DC} ( \operatorname{Proj}(\text {$\operatorname{I_{res}}$}) + \operatorname{I_{input}}), 
    \end{aligned}
\end{equation}
where $\operatorname{Proj}(\cdot)$ denotes the $1 \times 1$ convolutional layer and $\alpha$ is a learnable parameter.

\subsubsection{Data Consistency} 
Maintaining data consistency (DC) is an important step in MRI reconstruction, which prevents the network from modifying the observed $k$-space data. Following~\cite{duan2019vs}, we also include the DC operation in RDCN and EAM, which is an analytical solution naturally applicable to multi-coil data. Specifically, DC layer takes in four inputs: the input SENSE reconstruction $\mathbf{x_{in}}$, sub-sampled $k$-space data $\mathbf{y}$, the sub-sampling mask $M$ and the estimated sensitivity maps $S=\{S_i\}_{i=1}^{n_c}$. The overall DC process can be described as:
\begin{equation}
    \mathbf{x_{DC}} = \mathcal{R}(S, \mathcal{F^*}((\mathbf{1} - M)\odot \mathcal{F}(\mathcal{E}(\mathbf{x_{in}})) + M \odot y)).
\end{equation}

It is worth noting that both Reduce and Expand operation are used in our DC operation, where the Expand operation expands the SENSE reconstruction into $n_c$ coil images and the Reduce operation combines the coil images back into the SENSE reconstruction for the rest of the network.

\subsubsection{Loss Function} \label{loss_func}
In this work, we propose an edge-guided model that can be trained in an end-to-end manner. To ensure the accuracy and quality of the predicted edges, we apply supervised learning to the outputs of the EPN. Specifically, we design an edge-aware loss in this work, which is composed of two parts: image loss and edge loss. Image loss is computed over the difference between the Root Sum of Squares (RSS) of the final reconstructed MR image $\mathbf{x_{pred}}$ and the corresponding ground truth $\mathbf{x}$: 
\begin{equation}
    \operatorname{\mathcal{L}_{image}} = || \mathbf{x_{pred}} - \mathbf{x}||_1.
\end{equation}

On the other hand, to compute the edge loss, we first extract the ground truth edges using the \textbf{Sobel} operator. Then, we compute the difference between predicted edges $\mathbf{e_{pred}} = \{\mathbf{e_{pred}^{t}}\}_{t=1}^{N}$ and the the corresponding ground truth edges $\mathbf{e}$: 
\begin{equation}
    \operatorname{\mathcal{L}_{edge}} = \sum_{t=1}^{N}|| \mathbf{e_{pred}^t} - \mathbf{e}||_1,
\end{equation}
where $N$ denotes the number of cascaded blocks as shown in Fig.~\ref{arci}. For both losses, we adopt $L1$ norm due to its simplicity and effectiveness in producing sharp edges, which is also a common choice in other MRI reconstruction works like ~\cite{feng2022multi, wang2020deepcomplexmri}. Therefore, the total loss can be defined as:
\begin{equation}
    \operatorname{\mathcal{L}_{total}} = \operatorname{\mathcal{L}_{image}} + \beta \operatorname{\mathcal{L}_{edge}}.
\end{equation}
For simplicity, we set $\beta = 1$ in this work.

\section{Experiments Details}\label{sec:experiment}
\subsection{Datasets}

\subsubsection{Calgary-Campinas~\cite{souza2018open}}
Calgary-Campinas is a large-scale MR brain dataset jointly established by the Vascular Imaging Lab, University of Calgary and the Medical Image Computing Lab, University of Campinas. It provides both single- and multi-coil raw $k$-space data for evaluation. In this work, we first use the single-coil data to validate our model, which has been split into training (4,250 slices) and validation (1,700 slices) (testing data is not provided). The dataset is centrally cropped to $256 \times 256$. For multi-coil data, we use the provided T1-weighted, gradient-recalled echo, 1 mm isotropic sagittal dataset collected on a clinical MR scanner using a 12-coil array. The multi-coil dataset has been split into training (12,032 slices), validation (5,120 slices) and, testing (12,800 slices). Since there is no ground truth data for the testing data, we only use the training and validation data for evaluation. The dataset is centrally cropped to $218 \times 170$.

\subsubsection{fastMRI~\cite{zbontar2018fastmri}}
fastMRI is a large-scale MR dataset jointly established by Facebook AI Research and NYU Langone Health. It provides both knee and brain datasets for evaluation. In our work, we use the multi-coil knee dataset, which was acquired on three clinical 3T systems or one clinical 1.5T system using a 15-channel knee coil array. The dataset includes data from two pulse sequences, yielding coronal proton-density weighting with (PDFS) and without (PD) fat suppression. As is shown in Fig.~\ref{pd-pdfs-diff}, PD images usually contain more structural and prominent edge features than PDFS images, which suggests that it is more challenging to use edge guidance on PDFS datasets. Therefore, we explore the effectiveness of EAMRI on these two modalities. Following~\cite{feng2022multi}, for both PD and PDFS knee datasets, we separately filter out 227 volumes (8332 slices) for training and 24 volumes (1665 slices) for testing. The dataset is centrally cropped to $320 \times 320$.

\subsection{Implementation Details}\label{config}
Our model is implemented using PyTorch on an NVIDIA RTX A6000 GPU with 48GB of memory. We use Adam optimizer~\cite{kingma2014adam} with a learning rate of $5 * 10^{-4}$, the weight decay parameter was set as $10^{-7}$. For the single-coil task, we set mini-batch as 16, while for the multi-coil task, we set mini-batch as 8 for the Calgary dataset~\cite{souza2018open}, and 4 for the fastMRI dataset~\cite{zbontar2018fastmri}. All models were trained for 80 epochs in total. As for our network configuration, we use 4 RDCN for image de-aliasing (N = 4) and each RDCN contained 3 recursive DCB (M = 3). For the edge prediction network EPN, we use 3 cascaded MSRBs to extract multi-scale features. For EAM, we set the number of the edge attention heads as 4, and the number of the channels after the initial convolutions as 32.

For Calgary and fastMRI datasets, we use the sampling function provided by fastMRI\footnote{https://github.com/facebookresearch/fastMRI/} to generate Cartesian sampling mask with Acceleration Factor (AF) 4 and 6, respectively, which means that only 25\% and 16.7\% portion of the $k$-space data is used in the reconstruction process. For quantitative study, we use three evaluation metrics: peak signal-to-noise ratio (PSNR), structural similarity index (SSIM), and normalized mean square error (NMSE). 

\subsection{Baselines Models}
\subsubsection{Single-coil MRI Reconstruction}
Even though EAMRI is originally designed for multi-coil MRI reconstruction, it can be easily adapted to the single-coil MR datasets by removing the SME module and the Reduce operation at the beginning of the network. Therefore, we first evaluate it on the Calgary single-coil brain dataset. We compare our model with several lightweight and effective models for single-coil MRI reconstruction, including U-Net~\cite{hyun2018deep}, DCCNN~\cite{schlemper2017deep}, RDN~\cite{rdn}. Since there is no need to use the sensitivity map in the single-coil task, we simply delete the SME and Reduce operation in our model. For a fair comparison, we adjust the model size for all these models so that U-Net is 157K, DCCNN is 144K, RDN is 144K, MDR is 291K, and our EAMRI is 123K.

\subsubsection{Multi-coil MRI Reconstruction}
For the multi-coil task, we compare our model with U-Net~\cite{hyun2018deep}, DCCNN~\cite{schlemper2017deep}, E2EVarNet~\cite{sriram2020end}, VS-Net~\cite{duan2019vs} and RecurrentVarNet~\cite{yiasemis2022recurrent}. Among them, both U-Net and DCCNN are not originally proposed for multi-coil MRI reconstruction. Therefore we add the same SME module to their original architecture and feed them with the SENSE reconstruction. On the other hand, E2EVarnet, VS-Net, and RecurVarnet are SOTA deep unrolling methods that are derived from the optimization algorithms for the CS-based MRI reconstruction problem. All models except VS-Net were trained jointly with the same Sensitivity Map Estimation (SME) as our model. For VS-Net, we pre-compute the sensitivity map using BART~\cite{uecker2015berkeley} with parameters "bart ecalib -m1 -r26". For a fair comparison, we adjust the parameters for all these models to have comparable model sizes. After the adjustment, the number of parameters (in millions) for U-Net is 1.5M, DCCNN is 1.1M, E2EVarNet is 2.4M, RecurrentVarNet is 1.5M, VS-Net is 1.5M, and our EAMRI is 1.1M.

\section{Results}


\subsection{Results on Single-Coil MRI Reconstruction}
In Table~\ref{table:1}, we show the quantitative results for all model reconstructions on the Calgary single-coil brain dataset. 
According to the table, we can observe that EAMRI achieves the best results for the quantification metrics under acceleration factor 4 and requires fewer model parameters (123K). It is worth noting that even compared to the second best model DCCNN~\cite{schlemper2017deep}, PSNR is improved by 0.41dB. This effectively illustrates the excellence of EAMRI. 

\begin{figure*}[h!]
    \centering
    \begin{minipage}[c]{0.15\textwidth}
    \includegraphics[width=2.1cm]{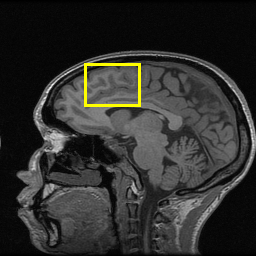}
    
    \includegraphics[width=2.1cm]{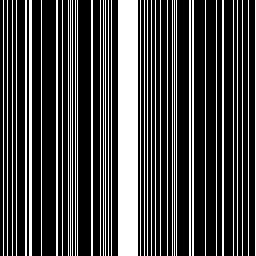}
    
    \includegraphics[width=2.1cm]{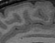}
    \centerline{GT/Mask}
    \end{minipage}
    \begin{minipage}[c]{0.15\textwidth}
    \includegraphics[width=2.1cm]{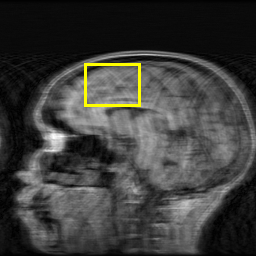}
    
    \includegraphics[width=2.1cm]{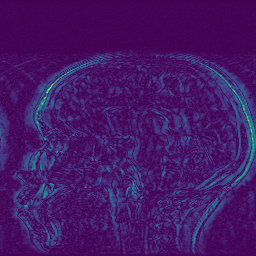}
    
    \includegraphics[width=2.1cm]{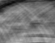}
    \centerline{Zero-Filled}
    \end{minipage}
    \begin{minipage}[c]{0.15\textwidth}
    \includegraphics[width=2.1cm]{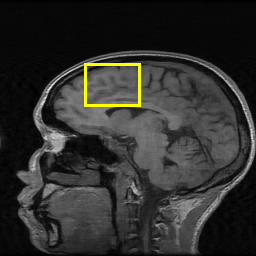}
    
    \includegraphics[width=2.1cm]{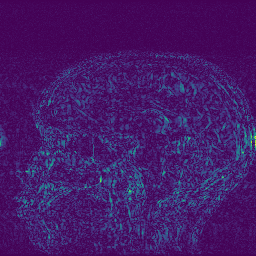}
    
    \includegraphics[width=2.1cm]{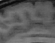}
    \centerline{U-Net~\cite{hyun2018deep}}
    \end{minipage}
    \begin{minipage}[c]{0.15\textwidth}
    \includegraphics[width=2.1cm]{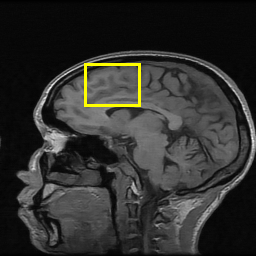}
    
    \includegraphics[width=2.1cm]{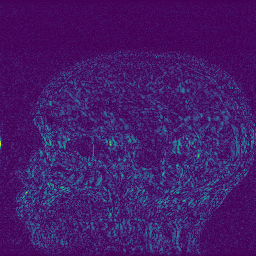}
    
    \includegraphics[width=2.1cm]{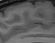}
    \centerline{DCCNN~\cite{schlemper2017deep}}
    \end{minipage}
    \begin{minipage}[c]{0.15\textwidth}
    \includegraphics[width=2.1cm]{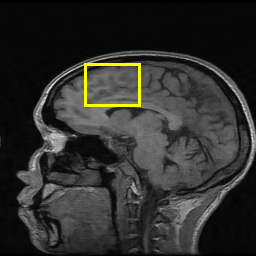}
    
    \includegraphics[width=2.1cm]{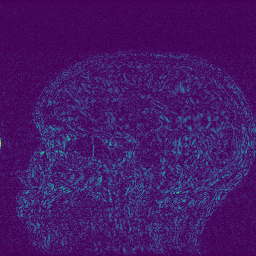}
    
    \includegraphics[width=2.1cm]{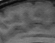}
    \centerline{RDN~\cite{rdn}}
    \end{minipage}
    \begin{minipage}[c]{0.15\textwidth}
    \includegraphics[width=2.1cm]{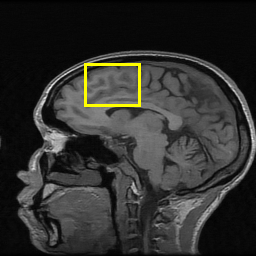}
    \includegraphics[width=2.1cm]{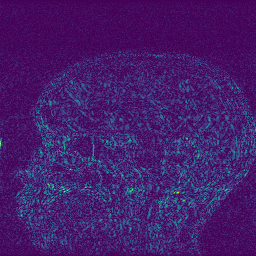}
    \includegraphics[width=2.1cm]{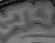}
    \centerline{EAMRI (Ours)}
    \end{minipage}  
    \caption{Qualitative results on Calgary~\cite{souza2018open} \textbf{Single-Coil} dataset with AF = 4. The first three rows are the reconstructed images, the heatmaps of the residual images and the zoomed-in images of the selected areas.}
\label{cc359_sc}
\end{figure*}

Moreover, in Fig.~\ref{cc359_sc}, we provide a visual comparison of the reconstruction results of these models. We can see that EAMRI has fewer bright spots in the heatmaps, which means less error between the EAMRI reconstructed image and the ground truth image. Meanwhile, according to the zoomed-in images of the selected areas, we can observe that our EAMRI can reconstruct more clean and accurate edges. This further validates the validity of EAMRI. Both the quantitative and the qualitative results for the single-coil MRI reconstruction demonstrate the effectiveness of EAMRI. 

\begin{table}[H]
\centering
\setlength{\tabcolsep}{8mm}
\caption{Quantitative results on Calgary~\cite{souza2018open} \textbf{Single-Coil} dataset. The best and the second-best results are highlighted in \textcolor{red}{red} and \textcolor{blue}{blue} color, respectively.}
\resizebox{.68\linewidth}{!}{
\begin{tabular}{|c|c|}
\hline
\multirow{2}{*}{Method} & AF = 4     \\ \cline{2-2} 
                        & \multicolumn{1}{l|}{PSNR (dB)$\uparrow$ / SSIM$\uparrow$ / NMSE$\downarrow$}  \\ \hline \hline
Zero-Filled (ZF)        & 27.36 / 0.8020 / 0.0517    \\ \cline{1-2} 
U-Net                   & 33.52 / 0.9188 / 0.0126    \\ \cline{1-2} 
DCCNN                   & \textcolor{blue}{35.51 / 0.9409 / 0.0082} \\ \cline{1-2}  
RDN                     & 34.42 / 0.9309 / 0.0104    \\ \cline{1-2} 
EAMRI (Ours)            & \textcolor{red}{35.92 / 0.9445 / 0.0075}  \\ \hline  
\end{tabular}}
\label{table:1}
\end{table}

\subsection{Results on Multi-Coil MRI Reconstruction}
In Table~\ref{table:2}, we provide the quantitative results of all the models over three multi-coil datasets: Calgary~\cite{souza2018open}, fastMRI PD~\cite{zbontar2018fastmri}, and fastMRI PDFS~\cite{zbontar2018fastmri}. According to the table, we clearly observe that EAMRI achieves the best performance over all these datasets under both AF 4 and AF 6. For the T1-weighted images (Calgary) and the PD images (fastMRI PD), whose edge features are more prominent and structural (see Fig.~\ref{pd-pdfs-diff}), EAMRI can achieve high-performance gain when compared with the second-best model. For example, when the AF = 4, PSNR is boosted by 0.64 over Calgary, and 1.26 over fastMRI PD. On the other hand, for the PDFS images under AF 4, PSNR is boosted by 0.22 when comparing EAMRI with the second-best model. This is because PDFS images usually contain more broken edges, which is more challenging to use the edge guidance mechanism. Fortunately, our EAMRI can still achieve better results, thanks to its well-designed network structure and edge priors.

\begin{figure}
    \centering
    \begin{minipage}[c]{0.135\textwidth}
    \includegraphics[width=2.0cm]{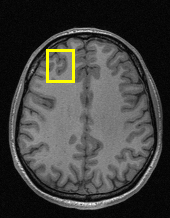}
    \includegraphics[width=2.0cm]{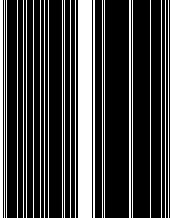}
    \includegraphics[width=2.0cm, height=1.1cm]{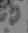}
    \centerline{(a)}
    \end{minipage}
    \begin{minipage}[c]{0.135\textwidth}
    \includegraphics[width=2.0cm]{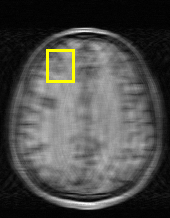}
    \includegraphics[width=2.0cm]{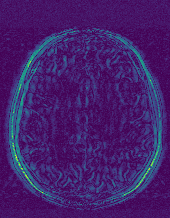}
    \includegraphics[width=2.0cm,height=1.1cm]{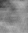}
    \centerline{(b)}
    \end{minipage}
    \begin{minipage}[c]{0.135\textwidth}
    \includegraphics[width=2.0cm]{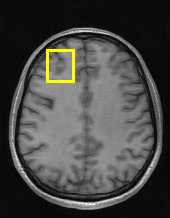}
    
    \includegraphics[width=2.0cm]{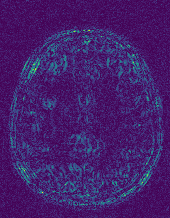}
    
    \includegraphics[width=2.0cm,height=1.1cm]{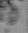}
    \centerline{(c)}
    \end{minipage}
    \begin{minipage}[c]{0.135\textwidth}
    \includegraphics[width=2.0cm]{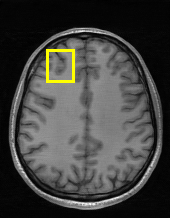}
    
    \includegraphics[width=2.0cm]{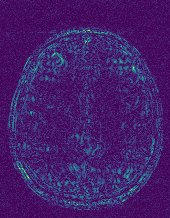}
    \includegraphics[width=2.0cm,height=1.1cm]{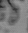}
    \centerline{(d)}
    \end{minipage}
    \begin{minipage}[c]{0.135\textwidth}
    \includegraphics[width=2.0cm]{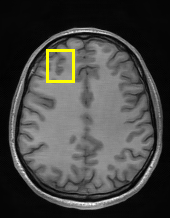}
    \includegraphics[width=2.0cm]{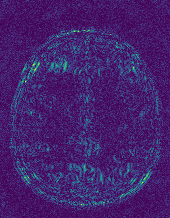}
    \includegraphics[width=2.0cm,height=1.1cm]{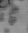}
    \centerline{(e)}
    \end{minipage}
    \begin{minipage}[c]{0.135\textwidth}
    \includegraphics[width=2.0cm]{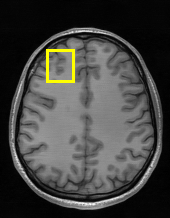}
    \includegraphics[width=2.0cm]{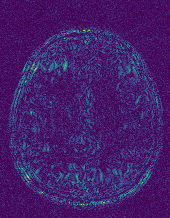}
    \includegraphics[width=2.0cm,height=1.1cm]{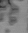}
    \centerline{(f)}
    \end{minipage}
    \begin{minipage}[c]{0.135\textwidth}
    \includegraphics[width=2.0cm]{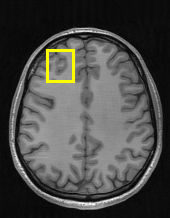}
    \includegraphics[width=2.0cm]{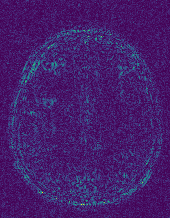}
    \includegraphics[width=2.0cm,height=1.1cm]{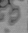}
    \centerline{(g)} 
    \end{minipage}  
    \caption{Qualitative results on Calgary~\cite{souza2018open} \textbf{T1-weighted Multi-Coil} dataset with AF = 4. From left to right: (a) Ground Truth, (b) Zero-Filled, (c) U-Net~\cite{hyun2018deep}, (d) E2EVarNet~\cite{sriram2020end}, (e) VS-Net~\cite{duan2019vs}, (f) RecurrentVarNet~\cite{yiasemis2022recurrent}, and (g) EAMRI (Ours)}
\label{cc359_mc_4}
\end{figure}

\begin{figure}
    \centering
    \begin{minipage}[c]{0.135\textwidth}
    \includegraphics[width=2.1cm]{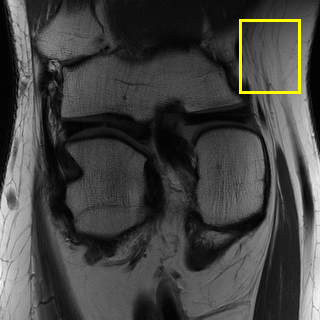}
    \includegraphics[width=2.1cm]{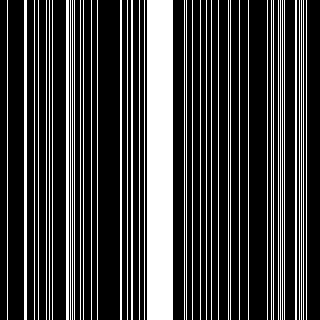}
    \includegraphics[width=2.1cm, height=1.3cm]{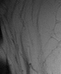}
    \centerline{(a)}
    \end{minipage}
    \begin{minipage}[c]{0.135\textwidth}
    \includegraphics[width=2.1cm]{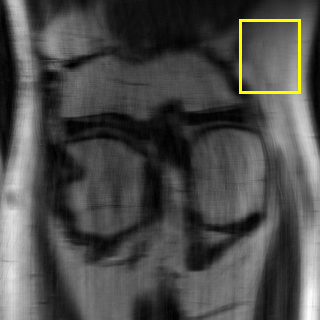}
    \includegraphics[width=2.1cm]{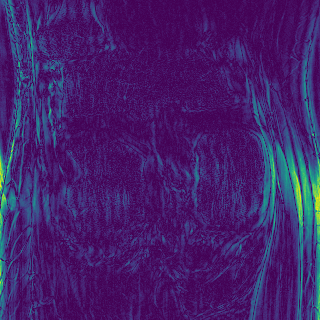}
    \includegraphics[width=2.1cm,height=1.3cm]{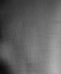}
    \centerline{(b)}
    \end{minipage}
    \begin{minipage}[c]{0.135\textwidth}
    \includegraphics[width=2.1cm]{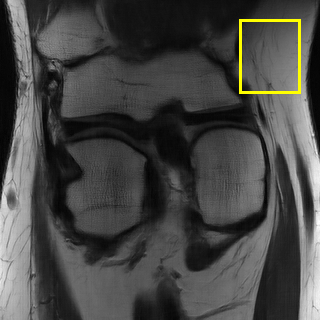}
    
    \includegraphics[width=2.1cm]{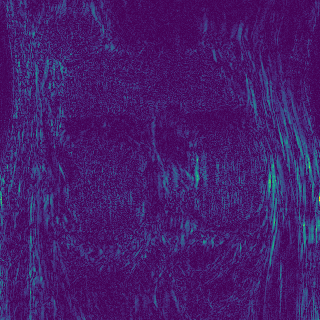}
    
    \includegraphics[width=2.1cm, height=1.3cm]{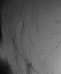}
    \centerline{(c)}
    \end{minipage}
    \begin{minipage}[c]{0.135\textwidth}
    \includegraphics[width=2.1cm]{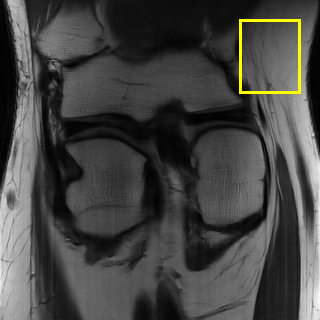}
    
    \includegraphics[width=2.1cm]{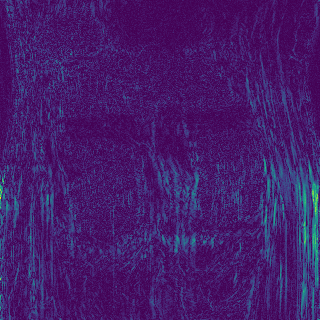}
    
    \includegraphics[width=2.1cm, height=1.3cm]{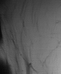}
    \centerline{(d)}
    \end{minipage}
    \begin{minipage}[c]{0.135\textwidth}
    \includegraphics[width=2.1cm]{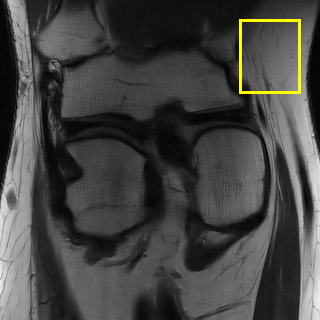}
    
    \includegraphics[width=2.1cm]{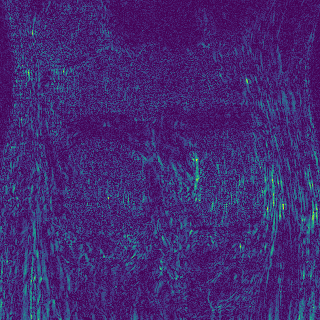}
    
    \includegraphics[width=2.1cm, height=1.3cm]{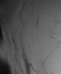}
    \centerline{(e)}
    \end{minipage}
    \begin{minipage}[c]{0.135\textwidth}
    \includegraphics[width=2.1cm]{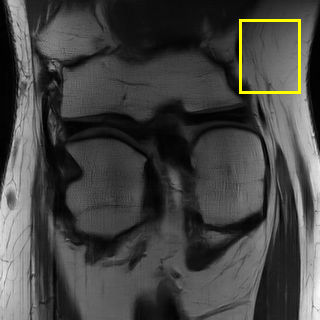}
    
    \includegraphics[width=2.1cm]{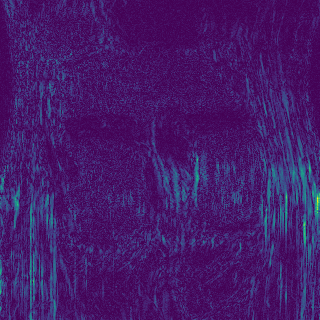}
    
    \includegraphics[width=2.1cm, height=1.3cm]{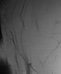}
    \centerline{(f)}
    \end{minipage}
    \begin{minipage}[c]{0.135\textwidth}
    \includegraphics[width=2.1cm]{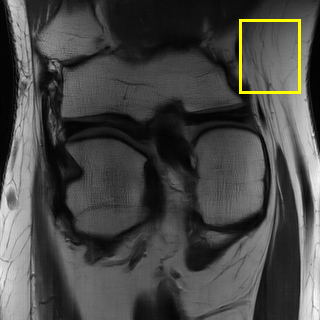}
    
    \includegraphics[width=2.1cm]{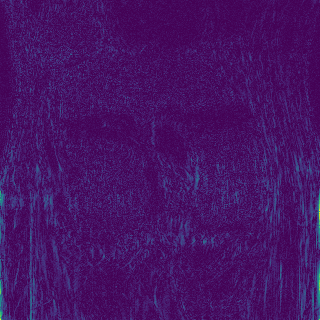}
    
    \includegraphics[width=2.1cm, height=1.3cm]{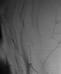}
    \centerline{(g)}
    \end{minipage}  

    \caption{Qualitative results on fastMRI~\cite{zbontar2018fastmri} \textbf{PD Multi-Coil} dataset with AF = 4. From left to right: (a) Ground Truth, (b) Zero-Filled, (c) U-Net~\cite{hyun2018deep}, (d) E2EVarNet~\cite{sriram2020end}, (e) VS-Net~\cite{duan2019vs}, (f) RecurrentVarNet~\cite{yiasemis2022recurrent}, and (g) EAMRI (Ours).}
\label{fastpd_mc_4}
\end{figure}

\begin{figure}
    \centering
    \begin{minipage}[c]{0.135\textwidth}
    \includegraphics[width=2.1cm]{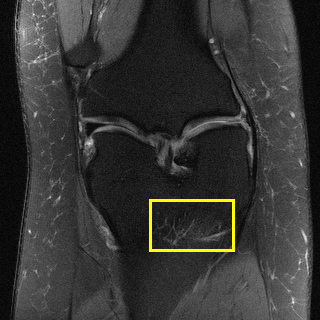}
    
    \includegraphics[width=2.1cm]{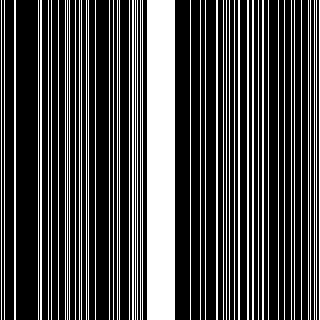}
    
    \includegraphics[width=2.1cm, height=1.3cm]{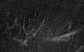}
    \centerline{(a)}
    \end{minipage}
    \begin{minipage}[c]{0.135\textwidth}
    \includegraphics[width=2.1cm]{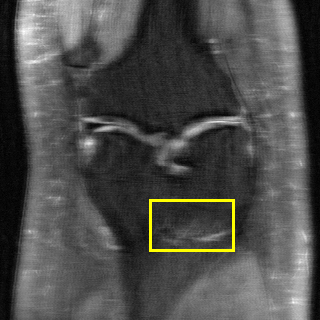}
    
    \includegraphics[width=2.1cm]{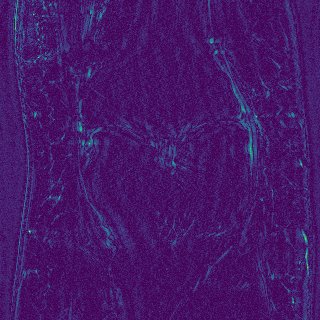}
    
    \includegraphics[width=2.1cm,height=1.3cm]{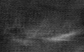}
    \centerline{(b)}
    \end{minipage}
    \begin{minipage}[c]{0.135\textwidth}
    \includegraphics[width=2.1cm]{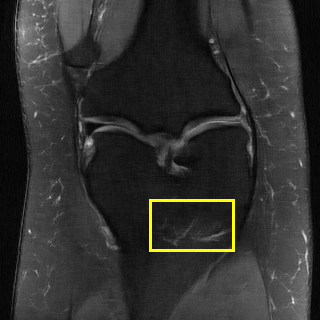}
    
    \includegraphics[width=2.1cm]{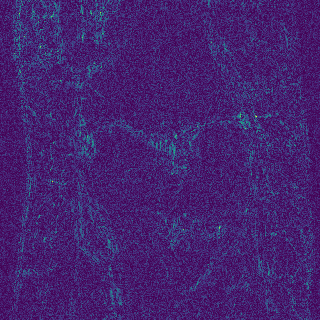}
    
    \includegraphics[width=2.1cm, height=1.3cm]{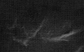}
    \centerline{(c)}
    \end{minipage}
    \begin{minipage}[c]{0.135\textwidth}
    \includegraphics[width=2.1cm]{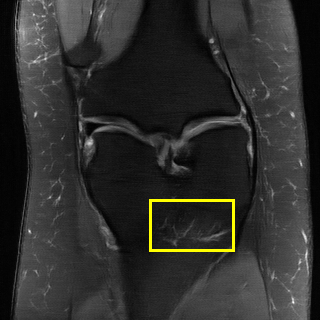}
    
    \includegraphics[width=2.1cm]{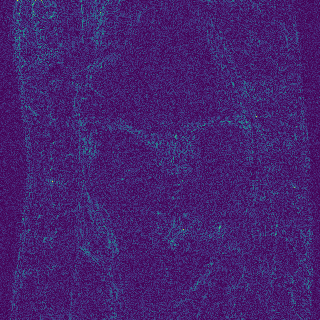}
    
    \includegraphics[width=2.1cm, height=1.3cm]{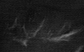}
    \centerline{(d)}
    \end{minipage}
    \begin{minipage}[c]{0.135\textwidth}
    \includegraphics[width=2.1cm]{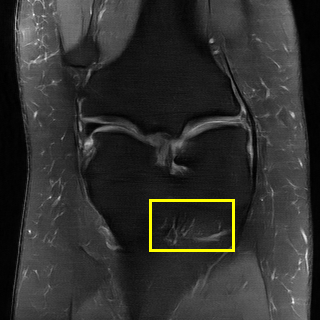}
    
    \includegraphics[width=2.1cm]{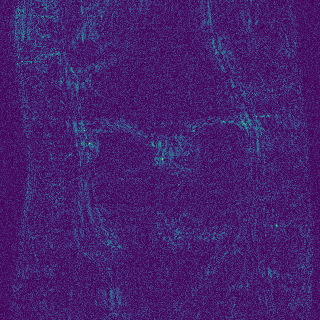}
    
    \includegraphics[width=2.1cm, height=1.3cm]{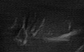}
    \centerline{(e)}
    \end{minipage}
    \begin{minipage}[c]{0.135\textwidth}
    \includegraphics[width=2.1cm]{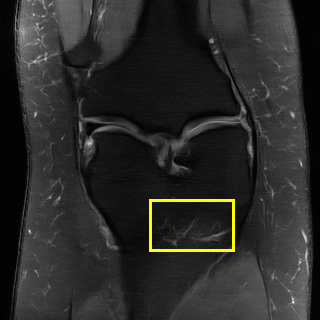}
    
    \includegraphics[width=2.1cm]{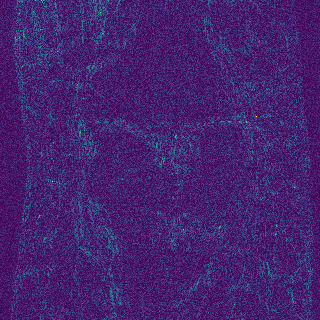}
    
    \includegraphics[width=2.1cm, height=1.3cm]{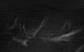}
    \centerline{(f)}
    \end{minipage}
    \begin{minipage}[c]{0.135\textwidth}
    \includegraphics[width=2.1cm]{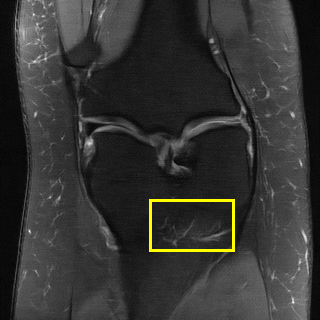}
    
    \includegraphics[width=2.1cm]{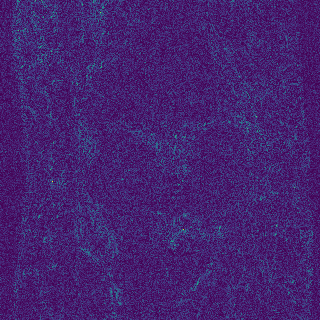}
    
    \includegraphics[width=2.1cm, height=1.3cm]{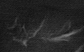}
    \centerline{(g)}
    \end{minipage}  
    
    \caption{Qualitative results on fastMRI~\cite{zbontar2018fastmri} \textbf{PDFS Multi-Coil} dataset with AF = 4. From left to right: (a) Ground Truth, (b) Zero-Filled, (c) U-Net~\cite{hyun2018deep}, (d) E2EVarNet~\cite{sriram2020end}, (e) VS-Net~\cite{duan2019vs}, (f) RecurrentVarNet~\cite{yiasemis2022recurrent}, and (g) EAMRI (Ours).}
\label{fastpdfs_mc_4}
\end{figure}

We also provide visual comparisons in Fig.~\ref{cc359_mc_4}-~\ref{fastpdfs_mc_4}. According to these images, we can see that EAMRI has fewer bright dots in the heatmaps and more clearer and accurate edges in the zoomed-in areas. The above quantitative and qualitative analysis fully demonstrate the effectiveness of the proposed EAMRI. Due to the page limit, more visual comparisons can be found at \url{https://github.com/MIVRC/EAMRI}.

\begin{table*}[h!]
\centering
\caption{PSNR/SSIM/NMSE comparisons with other lightweight MRI reconstruction models on Calgary~\cite{souza2018open} and fastMRI~\cite{zbontar2018fastmri} \textbf{Multi-Coil} dataset. The sampling rates for each dataset are 25\% (AF = 4) and 16.7\% (AF = 6), respectively. The best and the second-best results are highlighted in \textcolor{red}{red} and \textcolor{blue}{blue} color, respectively.}
\resizebox{\linewidth}{!}{
\begin{tabular}{|c|c|c|c|c|}
\hline
\multirow{2}{*}{Method} & \multirow{2}{*}{Data} & \multirow{2}{*}{Params} & AF = 4  & AF = 6    \\ \cline{4-5} 
                        &                               &                             & \multicolumn{1}{l|}{PSNR (dB)$\uparrow$ / SSIM$\uparrow$ / NMSE$\downarrow$} & \multicolumn{1}{l|}{PSNR (dB)$\uparrow$ / SSIM$\uparrow$ / NMSE$\downarrow$} \\ \hline \hline
                        
Zero-Filled (ZF)        & \multirow{7}{*}{Calgary} & - & 29.43 / 0.8070 / 0.0705   & 28.05 / 0.7607 / 0.0998 \\ \cline{1-1} \cline{3-5} 
U-Net       &           & 1.5M & 34.18 / 0.9186 / 0.0242     & 32.17 / 0.8855 / 0.0395 \\ \cline{1-1} \cline{3-5} 
E2EVarNet   &           & 2.4M & 34.28 / 0.9239 / 0.0239     & 32.13 / 0.8899 / 0.0407 \\ \cline{1-1} \cline{3-5} 
VS-Net      &           & 1.2M & \textcolor{blue}{35.62 / 0.9372 / 0.0178}     & \textcolor{blue}{33.17 / 0.9024 / 0.0318} \\ \cline{1-1} \cline{3-5} 
RecurrentVarNet &       & 1.5M & 35.29 / 0.9351 / 0.0188     & 32.85 / 0.8990 / 0.0335 \\ \cline{1-1} \cline{3-5} 
EAMRI (Ours)&           & 1.1M & \textcolor{red}{36.26 / 0.9445 / 0.0152}   & \textcolor{red}{33.87 / 0.9111 / 0.0276}      \\ \hline \hline 
Zero-Filled (ZF)        & \multirow{7}{*}{fastMRI PD} & -   & 31.60 / 0.8364 / 0.0219   & 29.17 / 0.7796 / 0.0382  \\ \cline{1-1} \cline{3-5} 
U-Net       &           & 1.5M & 38.40 / 0.9428 / 0.0046     & 35.57 / 0.9139 / 0.0089 \\ \cline{1-1} \cline{3-5} 
E2EVarNet   &           & 2.4M & \textcolor{blue}{39.55} / 0.9451 / 0.0036     & \textcolor{blue}{36.63 / 0.9161 / 0.0070} \\ \cline{1-1} \cline{3-5} 
VS-Net      &           & 1.2M & \textcolor{blue}{39.55 / 0.9524 / 0.0037}     & 36.22 / 0.9217 / 0.0080 \\ \cline{1-1} \cline{3-5} 
RecurrentVarNet &       & 1.5M & 39.11 / 0.9494 / 0.0040    & 35.85 / 0.9176 / 0.0084 \\ \cline{1-1} \cline{3-5} 
EAMRI (Ours)&           & 1.1M & \textcolor{red}{40.81 / 0.9591 / 0.0027}   & \textcolor{red}{37.98 / 0.9374 / 0.0052} \\ \hline  \hline
Zero-Filled (ZF)        & \multirow{7}{*}{fastMRI PDFS} & -   &  31.74 / 0.8011 / 0.0313 &  30.21 / 0.7486 / 0.0445  \\ \cline{1-1} \cline{3-5} 
U-Net       &           & 1.5M & 36.74 / 0.8816 / 0.0098    & 35.19 / 0.8605 / 0.0140 \\ \cline{1-1} \cline{3-5} 
E2EVarNet   &           & 2.4M & 37.06 / 0.8961 / 0.0092     & 35.62 / 0.8727 / 0.0128 \\ \cline{1-1} \cline{3-5} 
VS-Net      &           & 1.2M & 37.31 / 0.8975 / 0.0086     & 35.59 / 0.8721 / 0.0128 \\ \cline{1-1} \cline{3-5} 
RecurrentVarNet &       & 1.5M & \textcolor{blue}{37.55 / 0.9008 / 0.0081}    & \textcolor{blue}{35.75 / 0.8744 / 0.0123} \\ \cline{1-1} \cline{3-5} 
EAMRI (Ours)&           & 1.1M & \textcolor{red}{37.77 / 0.8981 / 0.0077}   & \textcolor{red}{36.29 / 0.8751 / 0.0109}       \\ \hline

\end{tabular}}
\label{table:2}
\vspace{-0.1cm}
\end{table*}

\section{ANALYSIS AND DISCUSSION}~\label{sec:ablation}
\subsection{Choices of Edge Detector}
To ensure effective edge guidance and accurate edge prediction, we impose supervision on the predicted edges. The ground truth edges can be extracted via multiple edge operators, like Sobel, Canny, and Prewitt. We select the two most representative edge detection operators: Sobel and Canny. Among them, Sobel is a simple and efficient operator that estimates the pixel-wise gradient magnitude by simply convolving the image with Sobel-x and y filters. Canny is a more complex and time-consuming operator that can produce smoother edges via non-maxima suppression and thresholding. We use these two operators to extract edge ground truth labels on clear MR images respectively and evaluate their performance on Calgary~\cite{souza2018open} multi-coil dataset with AF=4. Quantitative results are given in Table~\ref{operator}. According to the table, we can see that different edge detection operators will have an impact on the performance of the model. This reflects the validity and importance of the accuracy of edge priors in MRI reconstruction. Compared with the Canny operator, the use of the Sobel operator allows the model to achieve better results. It is worth noting that we believe that there are better edge detection operators that can extract better edge priors and further improve model performance. However, this is not the core of this work. Instead, the core of this work is to propose the edge guidance mechanism and verify the effectiveness of it. Therefore, in our model, we use the Sobel operator to extract ground truth edges due to its simplicity and effectiveness. 

\begin{table}[h!]
\centering
\setlength{\tabcolsep}{3mm}
\caption{Ablation studies for different edge detectors on Calgary~\cite{souza2018open} \textbf{Multi-Coil} dataset with AF = 4.}
\resizebox{0.65\linewidth}{!}{
\begin{tabular}{|c|c|c|}
\hline
\multirow{2}{*}{Category} & \multirow{2}{*}{Method}     & AF = 4 \\ \cline{3-3} 
                       & & \multicolumn{1}{l|}{\small PSNR (dB)$\uparrow$ / SSIM$\uparrow$ / NMSE$\downarrow$}  \\ \hline \hline

\multirow{2}{*}{Edge Detector}     & Canny         & 35.78 / 0.9360 / 0.0171    \\ \cline{2-3}
                                   & Sobel         & \textcolor{red}{36.26 / 0.9445 / 0.0152} \\ \hline
\end{tabular}}
\label{operator}
\end{table}

\subsection{Study on Edge Prediction Network}
\subsubsection{Number of MSRBs}
As shown in Fig.~\ref{RDB}, our proposed edge prediction network (EPN) utilizes multiple Multi-scale Residual Blocks (MSRBs) to extract multi-scale edge features. Therefore it is a natural question whether the number of MSRBs determines the quality of the predicted edges. To this end, we design three ablated models with different numbers of MSRBs. The quantitative results are given in Table~\ref{MSRB}. We can see that the performance of the model will improve as the number of MSRBs increases and the best performance is given when the number of MSRBs equal to 3. When the number of MSRBs increases to 5, the model performance will decrease. This is because too many MSRBs can lead to a sudden increase in the number of model parameters, which requires more training datas to fully train the model. Therefore, we use 3 MSRBs to build EPN to achieve the best balance between model 
size and performance.  

\begin{table}[h!]
\centering
\setlength{\tabcolsep}{4mm}
\caption{Ablation studies for different number of MSRB on Calgary~\cite{souza2018open} \textbf{Multi-Coil} dataset with AF = 4.}
\resizebox{0.65\linewidth}{!}{
\begin{tabular}{|c|c|c|}
\hline
\multirow{2}{*}{Block}     & \multirow{2}{*}{Params} & AF = 4 \\ \cline{3-3} 
                       & & \multicolumn{1}{l|}{\small PSNR (dB)$\uparrow$ / SSIM$\uparrow$ / NMSE$\downarrow$}  \\ \hline \hline
1 MSRB      & 1.106M & 36.13 / 0.9416 / 0.0158 \\
3 MSRBs     & 1.154M & \textcolor{red}{36.26 / 0.9445 / 0.0152} \\ 
5 MSRBs                & 1.201M & 36.10 / 0.9408 / 0.0160 \\ \hline
\end{tabular}}
\label{MSRB}
\end{table}

\subsubsection{Quality of the Predicted Edges} 
\begin{figure}
    \centering
    \begin{minipage}[c]{0.14\textwidth}
     \includegraphics[width=2.1cm]{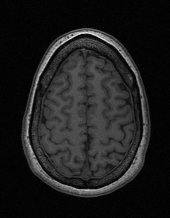}
     \includegraphics[width=2.1cm]{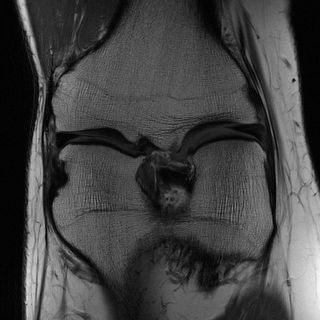}
     \includegraphics[width=2.1cm]{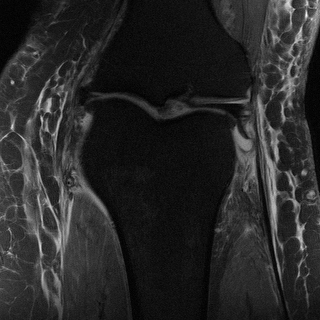}
     \centerline{\small (a)}
    \end{minipage}
     \begin{minipage}[c]{0.14\textwidth}
     \includegraphics[width=2.1cm]{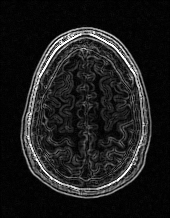}
     \includegraphics[width=2.1cm]{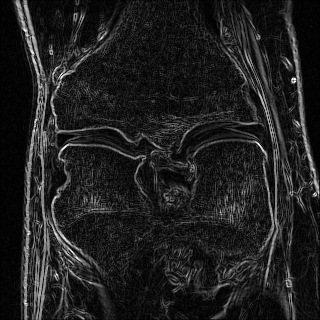}
     \includegraphics[width=2.1cm]{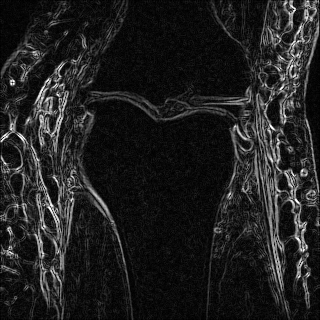}
     \centerline{\small (b)}
     \end{minipage}
     \begin{minipage}[c]{0.14\textwidth}
     \includegraphics[width=2.1cm]{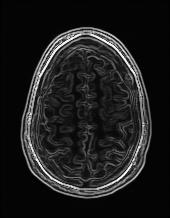}
     \includegraphics[width=2.1cm]{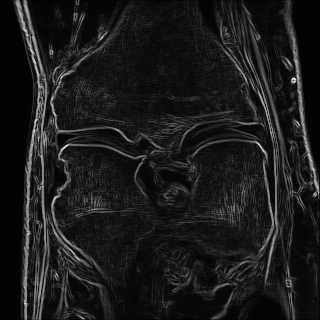}
     \includegraphics[width=2.1cm]{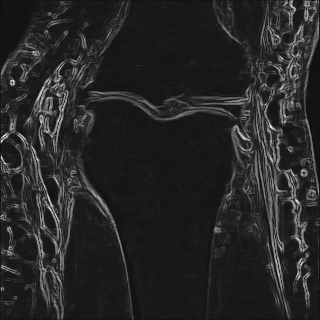}
     \centerline{\small (c)}
     \end{minipage}   
     \begin{minipage}[c]{0.14\textwidth}
     \includegraphics[width=2.1cm]{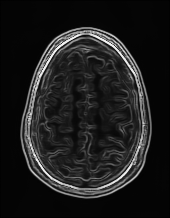}
     \includegraphics[width=2.1cm]{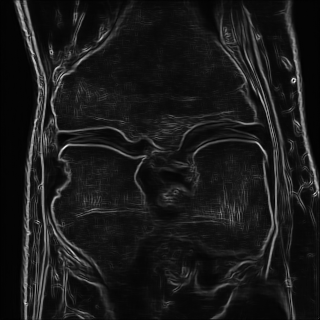}
     \includegraphics[width=2.1cm]{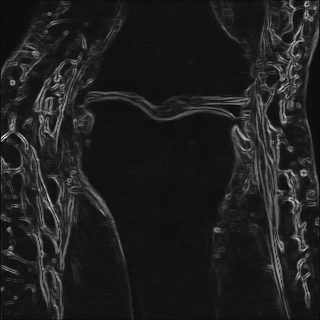}
     \centerline{\small (d)}
     \end{minipage}
\caption{\small Qualitative results of the predicted edges on Calgary~\cite{souza2018open} (top), fastMRI PD~\cite{zbontar2018fastmri} (middle), and fastMRI PDFS~\cite{zbontar2018fastmri} (bottom). From left to right: (a) GT, (b) GT edges extracted by Sobel, (c) Predicted edges with AF = 4, (d) Predicted edges with AF = 6.}
\label{edge_quality}
\end{figure}

EPN is utilized to provide edge priors for later reconstruction, so the quality of the predicted edges is very important. In Fig.~\ref{edge_quality}, we provide some qualitative results of the predicted edges of EPN on three multi-coil datasets. Among them, the GT edges are extracted using the Sobel operator. As can be seen from the images, our proposed EPN can predict an approximate contour for the overall subject and can reconstruct accurate edges close to the GT edges under two acceleration factors. This fully verifies the effectiveness and excellence of the proposed EPN.

\subsection{Study on Edge Attention Module}
In this work, we suggest using edge priors to guide MR image reconstruction. To this end, we propose an efficient guidance module, Edge Attention Module (EAM), to take full advantage of the predicted edge priors. To study the effectiveness of EAM, we design three different ablated models: (a) \textbf{M1}: a plain cascaded CNN \textit{without edge guidance}. This model adopts RDCN as basic image de-aliasing blocks without using EPN and EAM; (b) \textbf{M2}: an edge-guided CNN with the same architecture as our EAMRI, except that its EAM module is replaced by a \textit{simple edge guidance}, Concat + $1 \times 1$ Conv, to fuse the features of edges and images; (c) \textbf{M3}: EAMRI with \textit{shared weights EAM}. We compare these three models with our original EAMRI, and all these models have the same configuration regarding RDCN. The quantitative results are shown in Table~\ref{table:3}. 

When comparing M1 with the other models, we can see that the introduced edge guidance mechanism can improve the model performance, which verifies the validity of applying edge guidance in MRI reconstruction. Moreover, we can see that when comparing our final EAMRI with M2, PSNR is boosted by 0.54, and SSIM is boosted by 0.0086, with merely 0.4\% increase in model size, demonstrating that our designed EAM is a lightweight and effective module. On the other hand, when comparing M3 with our final EAMRI, we find that even though EAM with shared weights can save a few model weights, it will result in a performance downgrade. Therefore, we do not use the weight-sharing strategy in this module. Overall, this series of experiments further validate the importance of the edge guidance mechanism and also illustrates the feasibility and effectiveness of EAM.


\begin{table}[h!]
\centering
\setlength{\tabcolsep}{5mm}
\caption{Ablation studies of different modules on Calgary~\cite{souza2018open} \textbf{Multi-Coil} dataset with AF = 4.}
\resizebox{0.65\linewidth}{!}{
\begin{tabular}{|c|c|c|}
\hline
\multirow{2}{*}{Model}    & \multirow{2}{*}{Params} & AF = 4 \\ \cline{3-3} 
                       & & \multicolumn{1}{l|}{\small PSNR (dB)$\uparrow$ / SSIM$\uparrow$ / NMSE$\downarrow$}  \\ \hline \hline
M1     & 1.077M & 35.68 / 0.9343 / 0.0173 \\ 
M2  & 1.149M & 35.72 / 0.9359 / 0.0172 \\ 
M3  & 1.150M  & 36.18 / 0.9419 / 0.0155   \\
Final    & 1.154M &\textcolor{red}{36.26 / 0.9445 / 0.0152}    \\ \hline
\end{tabular}}
\label{table:3}
\end{table}

\section{Conclusion}\label{sec:conclusion}
In this work, we proposed a lightweight and accurate Edge Attention MRI Reconstruction Network (EAMRI) for accelerated MRI reconstruction. Specifically, we proposed an efficient Edge Prediction Network to directly predict image edges from the undersampled image and use them as external guidance for later reconstruction. Meanwhile, to fully take advantage of edge priors, we designed a novel Edge Attention Module, which can search for the best match between the image queries and the edge keys along the channel dimension. In this way image features can be globally activated by edge features, producing satisfactory reconstruction quality. Extensive experiments showed that our proposed EAMRI outperforms other methods with fewer parameters and can recover high-quality MR images with more accurate edges. This work provides promising guidelines for further research into multimodal MR imaging with transformers. Although our EAMRI provides excellent results on accelerated MR imaging, it still has some shortcomings. Like most supervised models, EAMRI inevitably suffers from performance decrease when it is tested on out-of-distribution medical data. As we know, current medical data is usually acquired by different scanners and protocols and from different medical centers. Therefore, as~\cite{nan2022data, yang2022unbox, lv2021transfer} suggest, we will consider combining data harmonization and transfer learning techniques with our current framework to enhance the model's robustness and generalizability.

\section*{Acknowledgments}
This work is supported in part by the National Key R\&D Program of China under Grant 2021YFE0203700 and 2021YFA1003004, in part by the Natural Science Foundation of Shanghai under Grand 23ZR1422200, in part by the Shanghai Sailing Program under Grant 23YF1412800, and in part by the NSFC/RGC N CUHK 415/19, Grant ITF MHP/038/20, Grant CRF 8730063, Grant RGC 14300219, 14302920, 14301121, and CUHK Direct Grant for Research.


\bibliographystyle{siam}

\end{document}